\documentclass[12pt]{article}

\usepackage{times}
\normalfont
\usepackage{mathptmx}
\usepackage{graphicx}

\usepackage{amsmath}

\allowdisplaybreaks

\title{
\begin{flushright}
{\normalsize{\tt hep-th/0703024}}\\
{\normalsize UT-Komaba/07-1}\\[10mm]
\end{flushright}
Holographic QCD and Pion Mass
}
\author{
Koji Hashimoto$^{\dagger a}$, 
Takayuki Hirayama$^{* b}$ and 
Akitsugu Miwa$^{\dagger c}$\\[10mm]
${}^\dagger$ {\it Institute of Physics, the University of Tokyo,}\\
\hspace{40mm}{\it Komaba 3-8-1, Tokyo 153-8902, Japan}\\
${}^*$ {\it Physikalisches Institut der Universitaet Bonn, }\\
\hspace{40mm}{\it Nussallee 12, 53115 Bonn, Germany}\\[10mm]
$^a$ {\normalsize E-mail: {\tt koji@hep1.c.u-tokyo.ac.jp}}\\
$^b$ {\normalsize E-mail: {\tt hirayama@th.physik.uni-bonn.de}}\\
$^c$ {\normalsize E-mail: {\tt akitsugu@hep1.c.u-tokyo.ac.jp}}\\[10mm]
}

\date{March, 2007}

\begin{document}

\baselineskip=16pt

\maketitle

\begin{center}{\bf abstract}\end{center}

\noindent
To realize massive pions, we study variations of the holographic model of
massless QCD using the D4/D8/$\overline{\rm D8}$ brane configuration
proposed by Sakai and Sugimoto. We propose deformations
which break the chiral symmetry explicitly and compute the mass of 
the pions and vector mesons. The observed value of the pion mass can be
obtained. We also argue a chiral perturbation corresponding to our
deformation.
\\[20mm]

\thispagestyle{empty}

\clearpage

\section{Introduction}

Holographic approach in analyses of QCD as an application of the
AdS/CFT correspondence \cite{Maldacena:1997re}
is not only a way to study strong coupling dynamics of QCD
in a perturbative way, but also giving a hope
to find a more comprehensive and deeper understanding of hadron physics
and holography~\cite{'tHooft:1993gx}
 itself. The developments in this research subject are 
still in the middle of the way to go beyond the large $N$ approximation
and to include dynamical quarks and various interactions. A lot remain to
be studied and to be revealed important and intriguing. 

Among many gravity models holographically dual to QCD, proposed so far
(see for example \cite{Karch:2002sh,Kruczenski:2003uq}), 
the Sakai-Sugimoto model \cite{Sakai:2004cn} is one of the most
successful models at present. An important feature of the model is that,
in terms of D-brane geometries, it explains the spontaneous chiral
symmetry breaking occurring at strong coupling in the real QCD. 
This is a typical example where D-branes in string theory 
provide a new interpretation of known physics --- any new
interpretation may help to create new techniques for analyzing
physical systems. Sakai and Sugimoto predicted various important 
phenomenological parameters associated with
hadron physics, such as 
vector/scalar meson spectra, interactions among them, 
chiral Lagrangian with calculable coefficients, skyrmions, and so on.
The comparison to the measured observable values is quite
successful, the values turned out to be within $20-30\%$
error~\cite{Sakai:2004cn}. 

The Sakai-Sugimoto model deals with massless QCD, thus the chiral
symmetry is unbroken at weak coupling. The spontaneous chiral symmetry
breaking is realized in terms of D-branes,
and massless pions appear as fluctuations on the D-branes.
In the real world, the quarks are massive and the chiral symmetry is
explicitly broken, thus the pions are 
pseudo Nambu-Goldstone bosons. To study the contribution of quark
masses to hadron physics, we have to introduce quark masses to the
holographic QCD model. Since the holographic QCD model describes hadron
physics, we aim in this paper to provide a D-brane construction to
introduce the pion mass in the Sakai-Sugimoto model, and to 
see how the hadron dynamics described in the massless model is modified
in the presence of the pion mass. A smooth limit to the Sakai-Sugimoto
model shows the existence of a corresponding chiral perturbation 
giving the pion mass.

In the Sakai-Sugimoto model, left-handed (or right-handed) quarks 
live on the intersection point of 
``gauge'' $N_c$ D4-branes and 
``flavor'' $N_f$ D8-branes (or $\overline{\rm D8}$-branes). 
The quark mass term mixes the left and the right, 
thus we need to bend the
D8-branes and the $\overline{\rm D8}$-branes and connect them {\it 
even at the weak coupling regime}, {\it i.e.} even as a D-brane
configuration in the flat spacetime background.  
This can be achieved by introducing a throat configuration of
D8-$\overline{\rm D8}$ branes~\cite{Callan:1997kz}
and placing D4-branes inside the throat, which will be studied in
section 2. Our result is that the throat surface
is located outside the near-horizon region of the D4-branes and
furthermore the pions are somehow still massless. Then we consider a
particular limit of this brane configuration, to make the D8-brane 
throat be flat and parallel to the
$N_c$ D4-branes, and put it inside the near horizon region. 
In this limit, the pion becomes massive as expected,
however it is too heavy (with no massless pion limit), and
so this brane configuration is not phenomenologically viable.
In section 3, we consider a different approach, which is introduction of 
a bound D4-brane charge on the D8- and the $\overline{\rm D8}$-branes
of the Sakai-Sugimoto model.
It is given by a Yang-Mills instanton on angular $S^4$ of the probe 
D8-brane worldvolume in the near-horizon background geometry.
This breaks the chiral symmetry explicitly, and with this, 
the value of
the pion mass is successfully tuned to be the realistic
one.  
We work out fluctuation analysis of the probe D8-brane
in the background geometry, following the computations in 
\cite{Sakai:2004cn}. We find that vector meson spectrum obtained in 
\cite{Sakai:2004cn} is not affected by the introduction of the pion
mass. We discuss a possible chiral perturbation
corresponding to this introduction of the D4-brane charge.

\section{Toward a holographic dual of QCD with massive quarks: some
 attempts}
\label{some attempts}

In the Sakai-Sugimoto model, the pure Yang-Mills part of the QCD 
is realized as a four-dimensional effective theory on 
$N_c$ D4-branes compactifying one spatial direction by an $S^1$ with
imposing the anti-periodic boundary condition for fermionic fields. They
introduce $N_f$ D8-branes and $N_f$ $\overline{\rm D8}$-branes which are
located at distinct points in the $S^1$ direction, and an open string
stretching from the D4-branes to the D8- or $\overline{\rm D8}$-branes
provides a chiral or anti-chiral fermion in four dimensions. The chiral
$U(N_f)_L\times U(N_f)_R$ symmetry is realized as a direct product of
gauge symmetries on the $N_f$ D8- and $N_f$ 
$\overline{\rm D8}$-branes. In the gravity dual description, where the
D8- and $\overline{\rm D8}$-branes can be treated as probes in the 
$N_f\ll N_c$ limit, the D8- and $\overline{\rm D8}$-branes have to
connect with each other smoothly in the near horizon geometry of the
corresponding D4-brane. This is interpreted as the
spontaneous chiral symmetry breaking in QCD, and they show the
existence of Nambu-Goldstone bosons associated with the spontaneous
chiral symmetry breaking, i.e. pions. However since the D4-branes
intersect with the D8- and $\overline{\rm D8}$-branes, the quark masses
are zero and then the pions are exactly massless.

\begin{figure}[t]
 \begin{center}
  \begin{minipage}{7cm}
   \begin{center}
    \includegraphics[width=6cm]{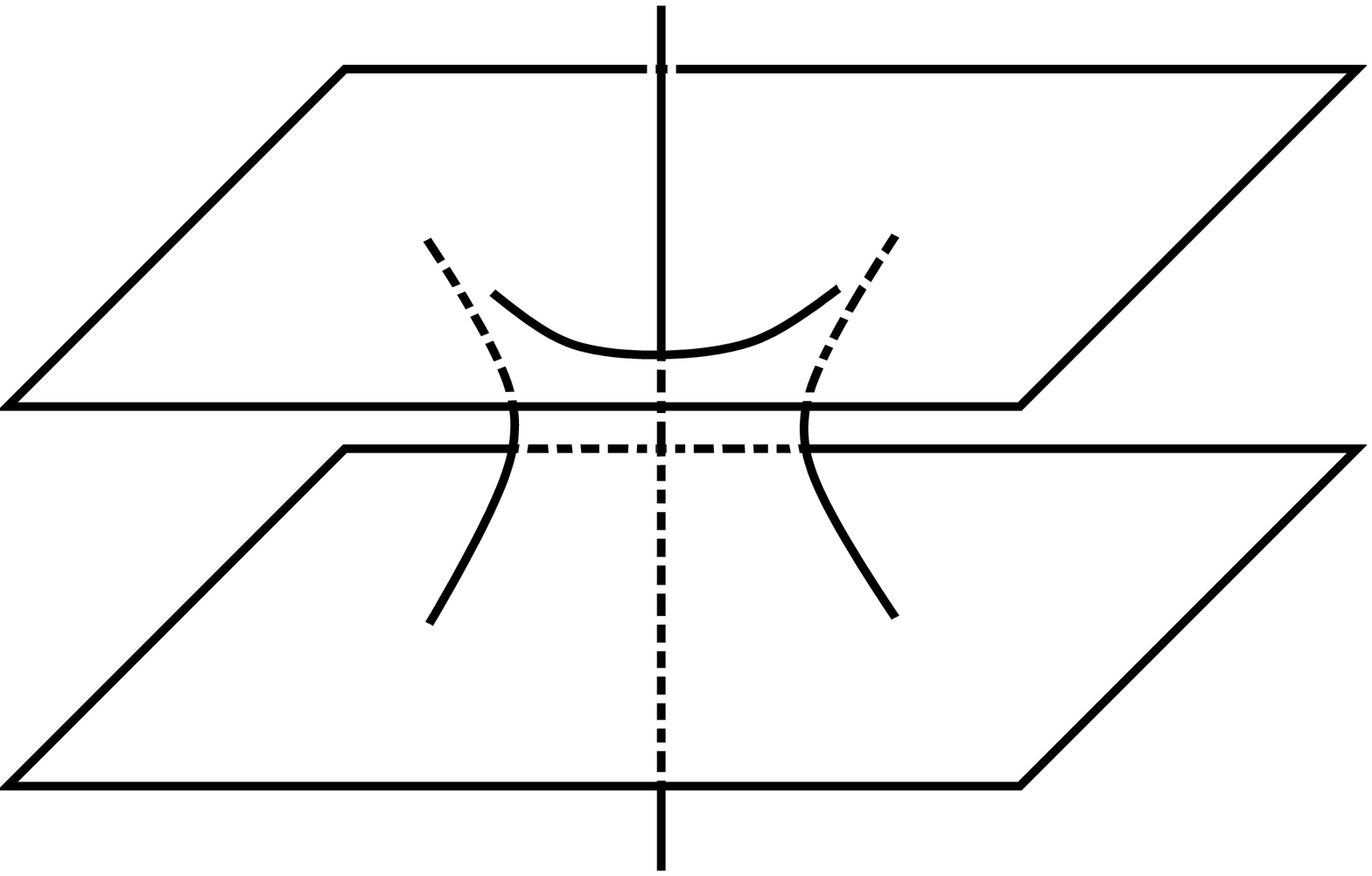}
    \put(-85,83){D4}
    \put(-45,83){D8}
    \caption{\footnotesize{The D4-branes are in the throat of the D8 
    $\overline{\rm D8}$.}}
    \label{fig4}
   \end{center}
  \end{minipage}
  \hspace{5mm}
  \begin{minipage}{9cm}
 \begin{center}
  \includegraphics[width=9cm]{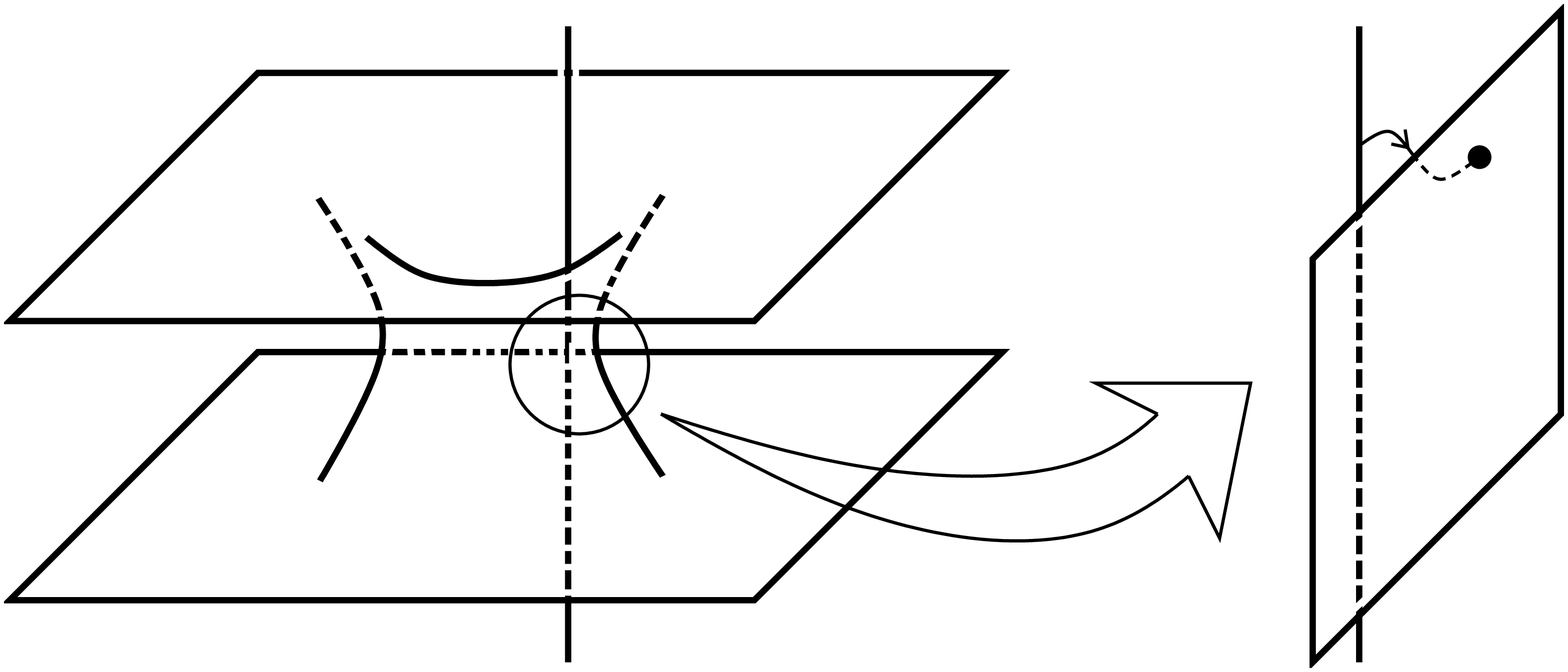}
  \caption{\footnotesize{The D4-brane location is shifted, and the 
  large throat limit is taken.}} 
  \label{fig6}
 \end{center}
  \end{minipage}
 \end{center}
\end{figure}

We expect that a quark becomes massive if we can deform the D8- and 
$\overline{\rm D8}$- brane configuration in such a way that they do not
intersect with the D4-branes in the flat spacetime. Although it is an
unstable configuration, it is known that there is such a static
configuration in which the parallel D8- and $\overline{\rm D8}$-branes 
are connected by a throat with the size almost equal to the asymptotic
distance between the D8- and $\overline{\rm D8}$- 
branes~\cite{Callan:1997kz}, see Figure \ref{fig4}.
We then place D4-branes inside the throat. Because the D8- and 
$\overline{\rm D8}$-branes no longer intersect with the D4-branes, and
also because the chiral symmetry is explicitly broken due to the fact
that the D8- and $\overline{\rm D8}$-branes are already connected, 
the masses of the quarks are expected to be non-zero and proportional to
the size of the throat radius, thus the pions become
pseudo Nambu-Goldstone bosons.\footnote{We notice that a tachyonic
mode appears in the scalar field on D8-branes, although this mode 
does not correspond to the pion.} 

The holographic dual gravity description for this D-brane throat should
be a new D8-brane probe configuration in the same D4-brane geometry.
In the flat spacetime, for a fixed distance between the D8- and the
$\overline{\rm D8}$-branes, there are two static configurations 
as we have already mentioned: the flat D8- and 
$\overline{\rm D8}$-branes, and the throat configuration. The former
corresponds to the original Sakai-Sugimoto model, while the latter is
what we are interested in. Thus, we expect that when the D4-branes are
replaced by their curved background geometry, we would have two probe
configurations, one for the former and the other for the latter. However,
the analysis in \cite{Sakai:2004cn} shows that, at least in the
near-horizon region, there is a unique probe configuration with the
asymptotic distance between the D8- and the $\overline{\rm D8}$-branes
fixed to be the anti-podal points in the $S^1$.\footnote{ 
In the paper~\cite{Aharony:2006da}, a one-parameter family of the
D8-brane probe configuration in the near-horizon geometry of the
D4-branes is obtained. The parameter is the asymptotic distance between
the D8- and the $\overline{\rm D8}$-branes in the $S^1$. The
D8- and $\overline{\rm D8}$-branes still intersect with the D4-branes in
the weak coupling picture ({\it i.e.} in the flat spacetime), thus
they have massless Nambu-Goldstone bosons in the strong coupling
picture.}

In order to find the missing solution, we study the probe D8-brane
configuration in the full D4-brane geometry without taking the low energy
limit, i.e. the near horizon limit. Now the boundary 
condition for the
D8-brane solution is imposed at the asymptotically flat region. Then we
find two probe D8-brane solutions for the same anti-podal boundary
condition: one extends into the near horizon geometry and the other does
not. The former solution corresponds to the D8-brane considered in
Sakai-Sugimoto model, and we identify the latter as what we are looking
for. Although the fields on D8-branes do not decouple with bulk gravity
modes and string massive modes (since we do not take the near horizon
limit), we may expect those modes do not break the chiral symmetry and
the pions are still massless if the quarks are massless. Conversely if
there are no massless pions, we may understand that the chiral symmetry
is explicitly broken due to the non-zero quark mass terms. We studied the
fluctuation on this D8-probe configuration to check if pions become
massive. On the contrary to our expectation from the D-brane picture in
the flat space, we find the pions are still massless (see more detail in
Appendix~\ref{hole}).

One possible explanation for this situation is that without taking the
near horizon limit, above picture just fails to capture the strong
dynamics of the dual QCD. However it may also be understood in the
following way. The amount of difference in energy from the D-brane
configuration in~\cite{Sakai:2004cn} is finite, which might imply that we
are looking at an excited state in the same theory (where quarks are
massless) and not a vacuum of a different theory. The operator
corresponding to the scalar field on the D8-brane, which determines the
probe D8-brane configuration, is not a quark bi-linear (the quark mass
term), but a four fermi term~\cite{Antonyan:2006vw} which can connect
left- and right-handed quarks without breaking the chiral symmetry. 
Nevertheless, this is still counter-intuitive, because
the open string stretching between the D8- and the D4-branes provides
a quark field and the energy of this open string 
is proportional to the distance between these two sets of
D-branes. In order to see this property more appropriately, we come to
consider a limit of the above D-brane configuration: taking the large
radius limit of the throat, and placing the D4-branes inside the 
throat not at the center, but at some fixed distance from the throat
surface. We ``magnify'' the throat region of the D8-branes, 
while keeping the distance to the D4-branes (see Figure \ref{fig6}).
After taking this limit the D8-branes are placed parallel to the
D4-branes. Then it is clear that the quarks are massive, since both the
D4- and the D8-branes are flat and separated by a non-zero
distance. Although there is no chiral symmetry because of the non-zero
masses for the quarks, we expect there are pseudo Nambu-Goldstone bosons
if the distance between the D4- and the D8-branes is much smaller than
the QCD scale.

However this limit causes a problem. The five dimensional theory before
compactification is a supersymmetric Yang-Mills theory with $N_f$
hypermultiplets. Therefore with the $S^1$ compactification with the
supersymmetry breaking boundary condition, the quark mass terms are at
least radiatively generated\footnote{
This system was first studied by the paper~\cite{Nastase:2003dd}.
The quark field can be periodic along the $S^1$ (the scalar partner is
then anti-periodic) in the field theoretical view point.
}
due to the lack of the chiral symmetry, even when the bare masses for the
quark fields are zero. Then we do not expect the pseudo Nambu-Goldstone
bosons. Despite of this, we might still expect that the mass of the pions 
is smaller than the masses of other mesons and that the mass spectrum is
close to the meson spectrum in the actual QCD. With this in mind, we
numerically computed the pion and the $\rho$ meson masses, and found that
the ratio is around $0.8$. This mass is clearly too large as the real
pions in QCD. We give more detailed calculations in Appendix~\ref{pp}.

\begin{figure}[t]
\begin{center}
\begin{minipage}{11cm}
\begin{center}
\includegraphics[width=6cm]{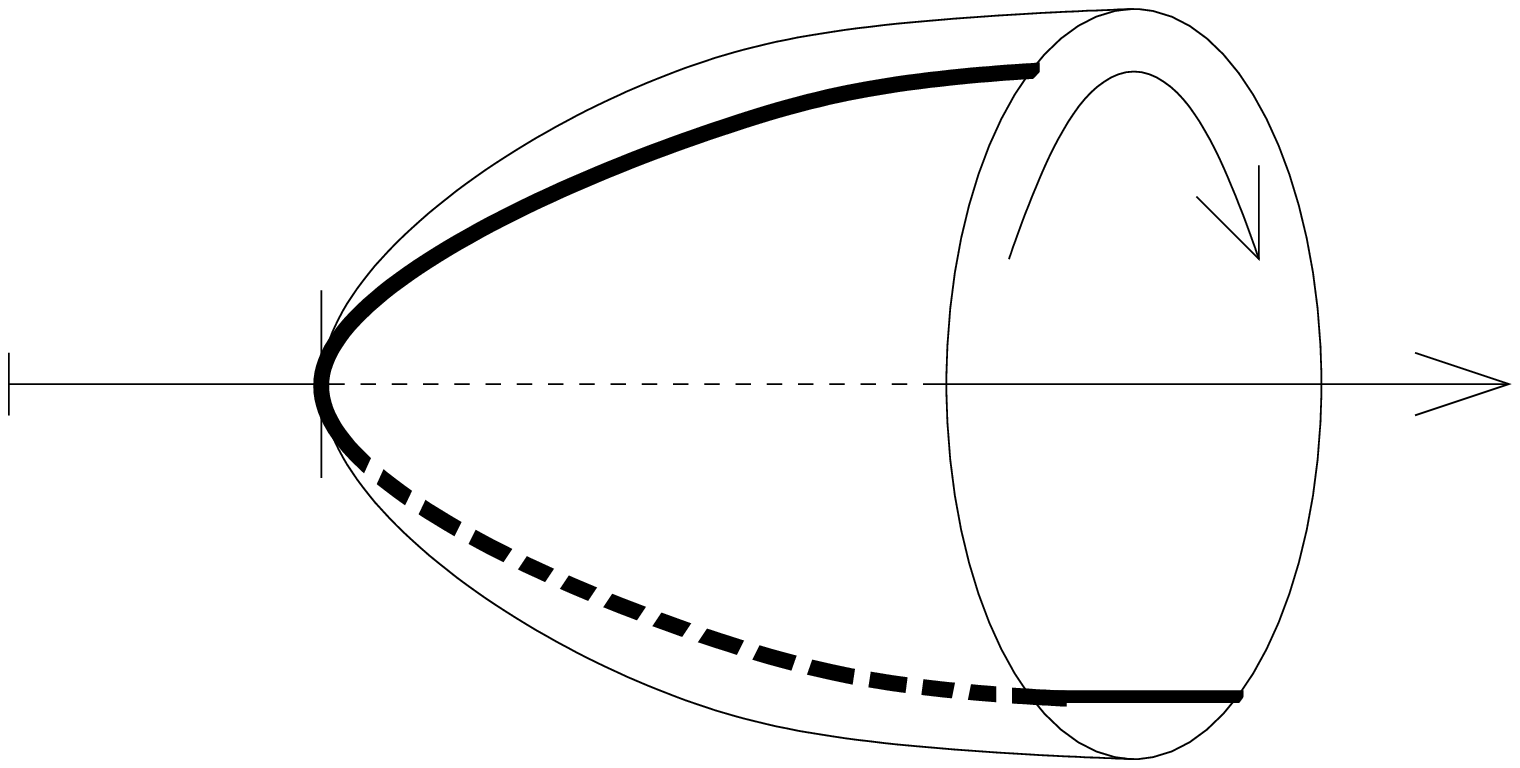}
\put(-190,30){$U=0$}
\put(-150,20){$U_{\rm KK}$}
\put(-10,25){$U$}
\put(-43,55){$\tau$}
\put(-90,60){D8}
\caption{\footnotesize{The D8-brane configuration localized at the
 anti-podal points on the $S^1$.}}
\label{fig2}
\end{center}
\end{minipage}
\end{center}
\end{figure}

\section{D4-brane charge and pion mass} 
\label{D4-brane charge and pion mass}

In the previous section, aiming to separate the D8-branes from the
D4-branes, we studied deformations of the probe D8-brane configuration. 
In this section we study a deformation expressed by a non-trivial 
background of gauge fields on the flavor D8-branes. In particular we
introduce a D4-brane charge on the D8-branes. One motivation for
introducing the D4-brane charge is that the conservation law of it
requires that the D8- and the $\overline {\rm D8}$-branes 
are connected to each other, even in the flat spacetime
background ({\it i.e.} in the weak coupling regime of the
QCD).\footnote{ This ``thin throat'' configuration of the 
D8-$\overline{\rm D8}$ is the one considered in 
\cite{Callan:1997kz} where a fundamental string charge is introduced
instead of our D4-brane charge.} From the view point of the effective
field theory on the D8-branes, this D4-brane is described by an instanton
in an angular $S^4$ of the worldvolume of the D8-branes
and thus breaks the chiral symmetry which is realized as a gauge symmetry
on the D8-branes. It may also be important that this deformation by
introducing the D4-brane charge on the $S^4$ extends infinitely along the
direction orthogonal to the color D4-branes since it implies the explicit
chiral symmetry breaking. From these perspectives, we expect that the
pions become massive and, as we will see, we indeed realize a
phenomenologically acceptable value of the pion mass.

The hadron physics in QCD is holographically captured by the
probe D8-branes in the D4-brane geometry in the Sakai-Sugimoto model.
In the following, we take the small $\alpha'$ limit in which the higher
derivative terms in the Dirac-Born-Infeld (DBI) action, as well as any
possible backreaction from the instanton to the shape of the brane, 
are negligible. Thus the system we analyze is the Yang-Mills theory on
the D8-brane probe, as considered by Sakai and Sugimoto. In particular we
concentrate on the case with $N_f=2$. In subsection \ref{instanton on D8}
we start with a brief review of a description of the probe D8-branes used
in \cite{Sakai:2004cn} and then introduce the instanton solution. After
that, in subsection \ref{meson spectrum}, we analyze spectra of
fluctuations of the gauge fields around this instanton background, which
corresponds to spectra of mesons appearing in the strong coupling regime
of QCD. We show that the instanton background successfully gives non-zero 
masses to the pions. Finally in subsection \ref{chiral perturbation} we
give an interpretation of the introduction of the instanton as a chiral
perturbation in QCD.

\subsection{Instanton on probe D8-brane} 
\label{instanton on D8}

We are interested in probe D8-branes extended into the near horizon
region of the non-extremal D4-brane background 
\cite{Witten:1998zw,Kruczenski:2003uq}
given by
\begin{align}
 & ds^2 = \bigg( {U \over R} \bigg)^{3/2} \big( dx_4^2 + f(U) 
 d\tau^2 \big) 
 + \bigg( {R \over U} \bigg)^{3/2} 
 \bigg( {d U^2 \over f(U)} + U^2 d \Omega_4^2 \bigg), \notag \\
 & e^\phi = \bigg( {U \over R} \bigg)^{3/4}, \quad
 F_4=\frac{2\pi N_c}{V_4}\epsilon_4,
 \quad f(U) = 1 - {U_{\rm KK}^3 \over U^3}.
 \label{kaba}
\end{align}
Here $d x_4^2 = \eta_{\mu \nu} dx^\mu dx^\nu$ is the line element of
the four-dimensional Minkowski spacetime and 
$d\Omega_4^2 = h_{ij} d\theta^i d\theta^j$ is that of a unit $S^4$.
The four form $\epsilon_4$ and the value $V_4$ are the volume form and
the volume of a unit $S^4$. The constant $g_s$ is the string coupling and
the parameter $R$ is related to the number $N_c$ of the color D4-branes.
The range of $U$ is limited by $U \geq U_{\rm KK}$ and the
$\tau$-direction is compactified on an $S^1$ with the periodicity 
$\delta \tau = 4 \pi R^{3/2}/3 U_{\rm KK}^{1/2}$. For fermions, we impose
the anti-periodic boundary condition along this $S^1$. With the
periodicity $\delta \tau$, $\tau-U$ plane is locally flat near 
$U=U_{\rm KK}$ without any conical singularity. The coordinates $x^\mu$
and $\tau$ are the ones along the D4-brane world volume.

The probe D8-brane worldvolume is on a plane defined by a constant
$\tau$. This is a solution since the metric does not depend on $\tau$
and the D8-branes are placed at the anti-podal points on the $S^1$,
see Figure \ref{fig2}. With use of a new coordinate $z$ defined by
$U^3 = U_z^3 \equiv U_{\rm KK}^3 + U_{\rm KK} z^2$,
the induced metric on the D8-branes can be written down as
\begin{equation}
 ds_{D8}^2   = g_{MN} d\sigma^M d\sigma^N
  =
  \bigg( {U_z \over R} \bigg)^{3/2} dx_4^2 
  + {4 \over 9} \bigg( {R \over U_z} \bigg)^{3/2} {U_{\rm KK} \over U_z} 
  dz^2  + \bigg( {R \over U_z} \bigg)^{3/2} U_z^2 d \Omega_4^2.
  \label{induced_metric}
\end{equation}
Here indices $M$ and $N$ run from $0$ to $8$.

As explained at the beginning of this section, we study the Yang-Mills
action on the above D8-brane solution
in the small $\alpha'$ limit:
\begin{equation}
 S_{D8} = T_{D8}(2\pi\alpha')^2 
  \int d^9\sigma e^{-\phi} \sqrt{-\det g} {1 \over 2} {\rm Tr} 
  F_{MN} F^{MN}.
 \label{S D8}
\end{equation}
Here $T_{D8}$ is the tension of a D8-brane and the field strength
$F_{MN}$ is defined by 
$F_{MN} \equiv \partial_M A_N - \partial_N A_M + [A_M , A_N]$.
(We use the convention in which the gauge fields $A_M=iA_M^aT^a$ are
anti-Hermitian matrices and the generators are normalized as 
${\rm Tr} T^a T^b = ( 1/2 ) \delta^{ab}$.) 
We introduce an instanton solution of this action and analyze its effects
on the meson spectrum. After the Kaluza-Klein (KK) reduction to four
dimensions, this action describes pions and vector mesons which are
realized as KK modes of the gauge fields. The instanton is introduced on
the $S^4$ and then the worldvolume of the induced D4-brane extends in the
non compact four dimensions $x^\mu$ and also in the $z$-direction. The
equations of motion for the gauge fields can be solved by the ansatz
\begin{equation}
 A_\mu =0, \quad A_z = 0, \quad A_i= A_i(\theta^j),
\end{equation}
with the self-dual conditions $F_{ij} = \ast F_{ij}$. Here $\ast$ is the
Hodge dual on a unit $S^4$. A solution of this self-dual equation can be
obtained from the SU(2) instanton solution on a flat 
${\bf R}^4$ \cite{Belavin:1975fg}:
\begin{align}
 A_a^{\rm inst} (X) = i {\epsilon_{abc} X^c \sigma_b - X^4 \sigma_a 
 \over \mu^2 + \rho^2},
 \quad (a=1,2,3)
 , \quad
 A_4^{\rm inst} (X) = i {X^a \sigma_a \over \mu^2 + \rho^2}, 
 \label{BPST}
\end{align}
by the stereographic projection:
\begin{align}
\begin{array}{ll}
 X^1  \equiv \cot \big( \theta^1 / 2 \big) \cos (\theta^2), 
& X^2  \equiv 
\cot \big( \theta^1 / 2 \big) \sin (\theta^2) \cos (\theta^3),\\
X^3  \equiv 
\cot \big( \theta^1 / 2 \big) \sin (\theta^2) \sin (\theta^3) 
 \cos (\theta^4),
& X^4  \equiv 
\cot \big( \theta^1 / 2 \big) \sin (\theta^2) \sin (\theta^3) 
 \sin (\theta^4).
\end{array}
\end{align}From the Jacobian of this transformation we have
$16d^4X=(\rho^2+1)^4d\Omega_4$. The coordinate systems $\{X^a,X^4\}$ 
and $\{\theta^i\}$ are the ones for a flat ${\bf R}^4$ and a unit $S^4$
respectively, and the coordinate $\rho \equiv |\vec X|$ is the radial
coordinate in the ${\bf R}^4$. The matrices $\{ \sigma_a \}$ are the
Pauli matrices. The parameter $\mu$ controls the size of the
instanton. The coordinates $X^a$, $X^4$ and the parameter $\mu$ are all
dimensionless in our notation. When we map the instanton solution
\eqref{BPST} with $\mu=1$ onto an $S^4$, we have a homogeneous instanton 
\cite{Constable:2001ag} which satisfies 
${\rm Tr} F_{ij} F^{ij} = {\rm constant}$.
In this case, the combined solution \eqref{induced_metric} and
\eqref{BPST} is not only the solution of the above Yang-Mills theory,
but also the solution of the original DBI action. For $\mu \neq 1$, the
instanton number density ${\rm Tr}F_{ij} F^{ij}$ becomes inhomogeneous
and dependent on $\theta^1$, or equivalently on $\rho$.

\subsection{Meson spectrum} \label{meson spectrum}

We now study the fluctuations of the gauge fields around the instanton
background and perform the KK reduction to four dimensions.
In \cite{Sakai:2004cn}, only the lowest modes with respect to the
$S^4$-coordinate dependence are considered, and mixing terms between  
$A_\mu$ (or $A_z$) with $A_i$ vanish. In our case with an instanton, the
mixing terms do not vanish. However in this subsection, we assume that
the mixing between fluctuations of $A_\mu$ or $A_z$ with that of $A_i$ is
small, and concentrate on the effect of the instanton on the
$A_\mu$-$A_z$ system. We discuss the mixing terms with $A_i$ in appendix 
\ref{mixing with A_i}.\footnote{
In appendix \ref{mixing with A_i}, we show that the lowest $S^4$ KK modes
of $A_i$ do not mix with those of $A_\mu$ nor $A_z$. Although we have not
evaluated the mixing terms including higher $S^4$ KK modes we expect that
these terms are suppressed because of the small overlap between the
wavefunctions.}

Plugging the induced metric (\ref{induced_metric}) into the action 
(\ref{S D8}) we obtain the four-dimensional Lagrangian:
\begin{align}
 S_{D8} &= \tilde T(2\pi\alpha')^2 \int d^4 x {\cal L},
 \\
 {\cal L} &= \int dz  { d\Omega_4 \over V_4}
 2 {\rm Tr}
 \bigg\{  
 {R^3 \over 4 U_z} \eta^{\mu \nu} \eta^{\rho \sigma} 
 F_{\mu \rho} F_{\nu \sigma}
 +
 {9 \over 8} {U_z^3 \over U_{\rm KK}} \eta^{\mu \nu} F_{\mu z} F_{\nu z}  
 \notag \\
 & \hspace{2.5cm}
 +
 {1 \over 2} \eta^{\mu \nu} h^{ij} D_i A_\mu D_j A_\nu 
 +
 {9 \over 8} {U_z^4 \over R^3 U_{\rm KK}} h^{ij} D_i A_z D_j A_z  
 + \textrm{( terms with $A_i$ )}
 \bigg\}. \label{action_mu_z}
\end{align}
Here $\tilde{T}=2R^{3/2}U_{KK}^{1/2}T_{D8}V_4/3$, and
$D_i A_\mu$ and $D_i A_z$ are defined by
$D_i A_M \equiv \partial_i A_M + \big[ A_i^{\rm inst}, A_M  \big]$
(where $M=\mu \,\,{\rm or} \,\, z$). We first perform the integration
over the $S^4$. Recalling that the instanton number density depends only
on $\rho = |\vec X|$, we assume that the wave functions of the lowest
$S^4$ KK modes of $A_\mu$ and $A_z$ depend only on $\rho$ accordingly.
Under this assumption, we can write these lowest modes as follows:
\begin{equation}
  A_\mu (x,z,X) = \tilde A_\mu (x,z) \zeta(\rho),\quad
   A_z   (x,z,X) = \tilde A_z (x,z) \zeta(\rho) .
 \label{Az=Az.phi.xi}
\end{equation}
We define $\zeta(\rho)$ by the eigenvalue equation 
(with the lowest eigenvalue $\epsilon^{2}$)
\begin{equation}
 -\partial_\rho \big( \rho^3 L \partial_\rho \zeta \big)
 + 8 L { \rho^5 \over ( \mu^2 + \rho^2 )^2 } \zeta
 =
 \rho^3 L^2 \epsilon^2 \zeta,
 \quad
 \left(L \equiv 4 (\rho^2 + 1)^{-2}\right),
 \label{eigen_eq_tlambda} 
\end{equation}
and the normalization condition 
$\int ( d\Omega_4 /V_4 )\zeta (\rho)^2 = 1$. 
Now the mass for $\tilde A_\mu$ ($\tilde A_z$ is similar) is generated 
from the third term on the right hand side in (\ref{action_mu_z}) as 
\begin{align}
 \int { d\Omega_4 \over V_4} \eta^{\mu \nu}h^{ij} {\rm Tr} 
 D_i A_\mu D_j A_\nu = -{1 \over 2} \epsilon^2
 \eta^{\mu \nu} \tilde A_\mu^a \tilde A_\nu^a, \label{A_mu mass}
\end{align}
where $a$ in the upper index denotes $SU(N_f)$ gauge index.
There are also higher modes on the $S^4$, which we do not consider in
this article. Performing the integration over the $S^4$ and 
keeping only terms quadratic in fluctuations, we obtain
\begin{align}
 {\cal L} = -\int dz 
 \bigg\{
 {R^3 \over 4 U_z} \eta^{\mu \nu} \eta^{\rho \sigma} 
 \tilde F_{\mu \rho}^a \tilde F_{\nu \sigma}^a
 +{9 \over 8}  {U^3_z \over U_{\rm KK}} \eta^{\mu \nu} 
               \tilde F_{\mu z}^a \tilde F_{\nu z}^a
 + {1 \over 2} \epsilon^2 \eta^{\mu \nu} 
               \tilde A_\mu^a \tilde A_\nu^a
 + {9 \over 8} {U_z^4 \over R^3 U_{\rm KK}} 
               \epsilon^2 \tilde A_z^a \tilde A_z^a  
 \bigg\}. \label{Lagrangian_dz}
\end{align}
Note that if the second term on the left hand side of
\eqref{eigen_eq_tlambda} do not exist, then $\zeta = {\rm constant}$ is
the normalizable eigen function with $\epsilon = 0$. If we consider an
instanton with large $\mu$, then the second term on the left hand side of
\eqref{eigen_eq_tlambda} is small compared to the other two terms for a
wide range of the variable $\rho$, and $\epsilon \sim \mu^{-1}$ as we
will see later. Thus for large $\mu$, the third and the fourth terms 
in \eqref{Lagrangian_dz} can be considered as small perturbations
added to the model considered by Sakai and Sugimoto.

Next we perform the $z$-integral. Since we treat the instanton with
$\mu^{-1} \ll 1$ as a small perturbation, we may determine mode functions
of $\tilde A_\mu$ and $\tilde A_z$ along the same line as
\cite{Sakai:2004cn}. Although with this choice of mode functions the 
KK modes are not diagonalized, we can diagonalize them after the
$z$-integral. 

Let us first put $\tilde A_z=0$ and determine the mode functions $\psi_m$
for $\tilde A_\mu(x,z) =
\sum_{m\geq 1}\tilde{A}_{\mu}^{(m)}(x)\psi_m(z)$. 
The equation for the mode functions which diagonalize the Lagrangian 
is easily obtained: 
\begin{equation}
 -\partial_z \big( K \partial_z \psi_m \big) 
 + {4 \over 9} \epsilon^2 U_{\rm KK}^{-2}\psi_m 
 = 
 \lambda_m U_{\rm KK}^{-2} K^{-1/3} \psi_m
 \quad \Bigg( K \equiv \bigg( { U_z \over U_{\rm KK} } \bigg)^3\Bigg) .
\label{eigen_eq_lambda}
\end{equation}
We choose the normalization condition
$\int dz K^{-1/3} \psi_m \psi_n=\delta_{mn}$.
As for the mode functions $\phi_m$ for
$\tilde A_z(x,z)=\sum_{m\geq 0}\tilde{A}_z^{(m)}(x)\phi_m(z)$, we choose 
\begin{equation}
 \phi_m \equiv \partial_z \psi_m \quad (m\geq1), \qquad \phi_0 \propto 
  {1 \over K},
\end{equation}
as in \cite{Sakai:2004cn}, and in the following we denote the lowest mode
as $\tilde{A}^{(0)}_z\equiv\varphi$. Using this mode expansion, we obtain
\begin{align}
 {\cal L} & =
 -\frac{9U_{\rm KK}^2}{8}\Bigg[K_{00}(\partial_\mu\varphi^a)^2
 +
 {4 \over 9} \epsilon^2 M_{\rm KK}^2 M_{00}(\varphi^a)^2 \Bigg]
 -\frac{R^3}{U_{\rm KK}}\sum_{m\geq 1}\Bigg[
 {1 \over 4} 
 \big(\tilde F_{\mu \nu}^{a,(m)}\big)^2
 + {1 \over 2} \lambda_m M_{\rm KK}^2 \big(\tilde A_\mu^{a,(m)}\big)^2
  \Bigg]
 \notag \\
 & \quad
 -\frac{9U_{\rm KK}^2}{8}\sum_{m,n\geq 1}\Bigg[
 K_{mn}\partial_\mu \tilde A_z^{a,(m)}\partial^\mu \tilde A_z^{a,(n)}
 +\frac{4}{9}\epsilon^2 M_{\rm KK}^2M_{mn}
 \tilde A_z^{a,(m)}\tilde A_z^{a,(n)}\Bigg]
 \notag \\
 &\quad
 +\frac{9}{4}U_{\rm KK}^2\sum_{m,n\geq 1}
 K_{mn}\tilde A_\mu^{a,(m)}\partial^{\mu}\tilde A_z^{a,(n)}
 -\epsilon^2U_{\rm KK}^2M_{\rm KK}^2
 \sum_{m\geq 1} M_{0 m}\varphi^a \tilde A_z^{a,(m)},
 \label{L_vec_diagonal}
\end{align}
with
\begin{equation}
 K_{mn} \equiv \int dz K \phi_m \phi_n, \quad
 M_{mn} \equiv \int dz K^{4/3} \phi_m \phi_n.
\end{equation}
Here the four-dimensional Lorentz indices are contracted by the flat
metric. The parameter $M_{\rm KK}$ is defined as 
$M_{\rm KK}=3U_{\rm KK}^{1/2}/2R^{3/2}$. With the above choice of
$\phi_m$, the mixing terms of $\varphi^a$ with $\tilde A_\mu^{a,(m)}$ 
vanish, i.e., $K_{0m}=0$. The third line denotes the mixing of the KK
modes $\tilde A_\mu^{a,(m)}$ with $\tilde A_z^{a,(m)}$ and that
of $\varphi^a$ with $\tilde A_z^{a,(m)}$. The mixing terms between
$\tilde A_\mu^{a,(m)}$ and $\tilde A_z^{a,(m)}$ can be absorbed by the
following field redefinition:
\begin{equation}
 \tilde B_\mu^{a,(m)}\equiv\tilde A_\mu^{a,(m)} -
  \frac{U_{\rm KK}^2}{\lambda_m}\sum_{n\geq 1}K_{mn}\partial_\mu 
  \tilde A_z^{a,(n)},
\end{equation}
and finally we have
\begin{align}
 {\cal L} & =
 -\frac{9U_{\rm KK}^2}{8}\Bigg[K_{00}(\partial_\mu\varphi^a)^2
 +
 {4 \over 9} \epsilon^2 M_{\rm KK}^2 M_{00}(\varphi^a)^2 \Bigg]
 -\frac{R^3}{U_{\rm KK}}\sum_{m\geq 1}\Bigg[
 {1 \over 4} 
 \big(\tilde F_{\mu \nu}^{a,(m)}\big)^2
 + {1 \over 2} \lambda_m M_{\rm KK}^2 \big(\tilde B_\mu^{a,(m)}\big)^2
  \Bigg]
 \notag \\
 &
 -\frac{9U_{\rm KK}^2}{8}\!\!\!\!\sum_{m,n\geq 1}\Bigg[
 K'_{mn}\partial_\mu \tilde A_z^{a,(m)}\partial^\mu \tilde A_z^{a,(n)}
 +\frac{4}{9}\epsilon^2 M_{\rm KK}^2M_{mn}\tilde A_z^{a,(m)}
 \tilde A_z^{a,(n)}\Bigg]
 -\epsilon^2U_{\rm KK}^2M_{\rm KK}^2
 \sum_{m\geq 1} M_{0n}\varphi^a \tilde A_z^{a,(n)},
\end{align}
with
\begin{equation}
 K'_{mn} = K_{mn}-\sum_{p\geq1}K_{mp}
           \frac{U_{\rm KK}^2}{\lambda_p}K_{pn}.
\end{equation}
Note that we still have mixing terms in the pion sector which we
diagonalize after calculating matrices $K_{mn}$ and $M_{mn}$. We can see
that, after the limit $\epsilon\rightarrow 0$ is taken, this Lagrangian
becomes the one for the pion and the vector mesons in
\cite{Sakai:2004cn}.

Adopting the same charge conjugation and parity assignment as in
\cite{Sakai:2004cn}, the lowest mass eigen mode of $A_z$,
i.e. $\varphi$ with a small mixing with $\tilde A_z^{(m)}$, should be
identified with the pion, and the modes $\tilde A_\mu^{(m)}$ correspond
to the vector mesons. The other modes in $A_z$ were ignored in
\cite{Sakai:2004cn} and we follow this rule as we think of our
deformation as a perturbation.\footnote{
In this paper we take the gauge fixing condition in which the modes of
$A_i$ are eaten by the massive modes of $A_\mu$, on the other hand in
\cite{Sakai:2004cn}, the modes of $A_z$ are eaten by the massive modes of
$A_\mu$.
}
We expect that the mixing of $\varphi$ with each mode $\tilde A_z^{(m)}$ 
gets smaller as $m$ becomes larger. We compute the relevant matrices
$K_{mn}$ and $M_{mn}$ up to level $m=5$ and diagonalize the system.
In Table~\ref{vector meson} we summarize our final results of the pion
and the vector meson masses. We have used the actual observed
$\rho$-meson mass $m_\rho=776$ (MeV) as our input parameter.
\begin{table}
\begin{center}
\begin{tabular}{lc|c|cccccc}
   \hline
    $\mu^{-1}$ &&0&0.02 & 0.05 & 1/13.0 & 0.1 & 0.2 & 1.0\\
   \hline
   $\epsilon$ &&0& 0.0488 & 0.120 & 0.180 & 0.230 & 0.423 & 1.41 \\
   \hline
   \hline
   $m_{\pi^\pm,\pi^0}$ &(140,135)
 &0& 36.4 & 88.7 & 132 & 167 & 285 & 624\\
   \hline
   $m_\rho$ &(776) & (776)& (776) & (776) & (776) & (776) & (776) & (776)\\
   \hline
   $m_{a_1}$ &(1230)& 1189 & 1188 & 1186 & 1183 & 1179 & 1162 & 1046 \\
   \hline
   $m_{\rho'}$ &(1465)& 1607 & 1607 & 1603 & 1596 & 1589 & 1550 & 1308 \\
   \hline
\end{tabular}
 \caption{\footnotesize{
 The result of the numerical calculation of the meson spectrum. 
 The unit is MeV. The results for $\mu^{-1}=0$ corresponds to
 the massless QCD~\cite{Sakai:2004cn}.
 We have calculated matrices $K_{mn}$ and $M_{mn}$ up to $m=n=5$.
 We use $\rho$ meson mass $m_\rho = 776 ({\rm MeV})$ as our input 
 parameter.
 }
 \label{vector meson}}
\end{center}
\end{table}From Table \ref{vector meson}, we understand that $\epsilon$
is of the
order $\mu^{-1}$ ($\mu$ is the instanton size in ${\bf R}^4$), and 
that turning on a small $\mu^{-1}$ successfully generates the small
mass for the pions. Around $\mu=13$, the ratio $m_\rho / m_\pi$ comes 
close to the ratio of the actual observed masses. On the other hand, the
spectrum of the other vector mesons is not much affected compared with 
that of Sakai and Sugimoto \cite{Sakai:2004cn}.

Although we have succeeded in deriving the correct value of the pion mass,
it does not necessarily mean that we have succeeded in adding non-zero
quark masses to QCD. It is worth noting that the instanton did not
introduce a mass term for the possible Nambu-Goldstone boson
associated with the spontaneous symmetry breaking of
the axial $U(1)$ part of the $U(2)_L \times U(2)_R$ chiral symmetry.
This point may suggest that our deformation differs from adding the
quark masses. In the next subsection, we consider what kind of
perturbation corresponds to the introduction of the 
instanton from the view point of chiral perturbation. There we will show
that considering the property of the $A_i^{\rm inst}$ under the chiral
symmetry transformation, a four fermi coupling is more natural than the
quark mass terms as a possible lowest perturbation to the massless QCD.
However one may still wonder whether there is a relation between
the length of a stretched string and the quark mass. We will discuss this
point further in the appendix \ref{Small throat solution} by considering
the distance between the D4-branes and connected D8-$\overline {\rm D8}$
branes in the weak coupling regime, i.e., in the flat spacetime
background.
Though in this paper we worked for the case of $N_f=2$, 
for $N_f>2$ one can
introduce multi-instanton configurations on $S^4$ to reproduce 
the observed structure of mass spectrum of $\pi/K$.

\subsection{D4-branes as a chiral perturbation} 
\label{chiral perturbation}

Since the pion mass derived in the previous subsection is well below the
QCD scale, we should be able to understand the introduction of an
instanton in terms of a chiral perturbation. In this subsection we follow
the discussion by Sakai and Sugimoto on the derivation of the pion
effective action and study the chiral perturbation. As studied in the
previous subsection, since the instanton does not affect the pion
associated with the axial $U(1)$ breaking, we study a chiral perturbation
concerning only the pions associated with the chiral $SU(N_f)$ breaking.

We define a group element of the chiral symmetry transformation 
following the paper \cite{Sakai:2004cn}. In the previous subsection, we
have considered fluctuations of the gauge fields which vanish at 
$z \to \pm \infty$. The gauge transformation with an element 
$g(x,z,\theta) \in SU(N_f)$ which approaches a constant in the limit 
$z \to \pm \infty$ does not change this asymptotic behavior of the gauge
fields and is a transformation of residual gauge symmetry. The constant
of this transformation is identified as an element of the chiral symmetry
transformation $(g_+,g_-) \in SU(N_f)_L \times SU(N_f)_R$, 
$g_\pm \equiv \lim_{z\to \pm \infty} g(x,z,\theta)$.
If we introduce the following field $U(x,\theta)$:
\begin{equation}
 U(x,\theta) 
 \equiv
 {\rm P}\exp\bigg\{ - \int_{-\infty}^\infty dz A_z(x,z,\theta)  \bigg\},
 \label{PION}
\end{equation}
it is transformed as $U \to g_+ U g_-^{-1}$ under the chiral
transformation. For our later convenience we further introduce
\begin{equation}
 \xi_\pm^{-1}(x,\theta) \equiv
  {\rm  P}\exp\left\{-\int_0^{\pm\infty} dz' A_z(x,z',\theta)\right\},
  \hspace{3ex}
  A_i^\pm (\theta) \equiv A_i^{\rm inst} (x,z=\pm\infty,\theta) 
  = A_i^{\rm inst}(\theta).
\end{equation}
Then we have an expression 
$ U(x,\theta) = \xi_+^{-1}(x,\theta) \xi_- (x,\theta) $
and the transformation rules of $\xi_\pm$  and $A_i^{\pm}$ under the 
residual gauge symmetry above are as follows:
\begin{equation}
 \xi_+ \to h(x,\theta) \xi_+ g_+^{-1}, \quad
 \xi_- \to h(x,\theta) \xi_- g_-^{-1}, \quad
 A_i^+ \to g_+ A_i^+ g_+^{-1}, \quad 
 A_i^- \to g_- A_i^- g_-^{-1}, \label{baka}
\end{equation}
with $h(x,\theta) \equiv g(x,z=0,\theta)$. Note that we can take
$h(x,\theta)$ and $g_\pm$ as independent group elements since $g_\pm$
does not fix the gauge transformation parameter at finite $z$. Thus
we can choose the gauge $\xi_-=1$ (and therefore $\xi_+^{-1}=U$)
by using the degree of freedom of $h(x,\theta)$.\footnote{
Note that none of the group elements $g_\pm$, $h(x,\theta)$ nor 
$\xi_\pm(x,\theta)$ have nontrivial winding on the $S^4$.
}

We move to the $A_z=0$ gauge using the gauge transformation by the
group element ${\rm P} \! \exp( - \int_0^z A_z dz')$. Then the asymptotic
behavior of the gauge fields at $z\rightarrow\pm\infty$ now becomes
\begin{eqnarray}
 A_\mu(x,z,\theta) \rightarrow \xi_\pm(x,\theta)\partial_\mu
  \xi_\pm^{-1}(x,\theta), 
  \hspace{3ex}
 A_i(x,z,\theta) \rightarrow \xi_\pm(x,\theta) 
 (A^\pm_i(\theta)+\partial_i)
 \xi_\pm^{-1}(x,\theta), \label{asymp Amu Ai}
\end{eqnarray}
respectively. From this behavior and the above mentioned gauge choice,
$\xi_-=1$ and $\xi_+^{-1} = U$, we can expand the fluctuations with
respect to the mode functions of $z$-direction as follows:
\begin{align}
 &A_{\mu}(x,z,\theta)= U^{-1} (x,\theta)\partial_\mu U(x,\theta)
  \psi_+(z) 
 + \textrm{higher modes}
 ,
  \label{A_mu U}
  \\
 &A_i(x,z,\theta)=U^{-1}(x,\theta)(A_i^+(\theta)+\partial_i)
  U(x,\theta) \widetilde \psi_+(z)
  +  A_i^-(\theta) \widetilde \psi_-(z) 
 + \textrm{higher modes}
 .
 \label{A_i U}
\end{align}
Here, since we treat the instanton as a perturbation, we take $\psi_\pm$
and $\widetilde \psi_\pm$ as the zero modes for the case of the absence
of the instanton background. The explicit forms of these functions can be
read from the action:
\begin{equation}
 \psi_\pm(z) = \frac{1}{2} \bigg( 1 \pm { C_{-1}(z) 
  \over C_{-1}(\infty) } \bigg), \quad
  \widetilde \psi_\pm(z) = \frac{1}{2} \bigg( 1 \pm { C_{-4/3}(z) 
  \over C_{-4/3}(\infty) } \bigg), \quad
  C_n (z) = \int_0^z dz K^n.
\end{equation}
In the following, we will neglect the higher modes in the $z$-space 
in \eqref{A_mu U} and \eqref{A_i U}, since we are interested only in
the pion fields. Using \eqref{A_mu U} and \eqref{A_i U}, the field
strengths are computed as
\begin{align}
 F_{\mu\nu}&= [U^{-1}\partial_\mu U,U^{-1}\partial_\nu U]
  \psi_+(\psi_+-1),
  \\
 F_{z\mu}&= U^{-1}\partial_\mu U \partial_z \psi_+,
  \\
 F_{\mu i}&=  [U^{-1}\partial_\mu U, U^{-1} {\cal D}_i U ]
  \psi_+ \widetilde \psi_+ 
  + U^{-1} D_i^+ \big( \partial_\mu U U^{-1} \big) U \widetilde \psi_+
  - D_i^- \big( U^{-1} \partial_\mu U \big) \psi_+,
  \\
 F_{zi}&= U^{-1} {\cal D}_i U  \partial_z \tilde \psi_+
 \label{F_zi},
 \\
 F_{ij}&=
 [U^{-1} {\cal D}_i U,U^{-1} {\cal D}_j U] 
 \widetilde \psi_+ (\widetilde \psi_+ - 1)
 +
 U^{-1} F_{ij}^+ U \widetilde \psi_+ 
 +
 F_{ij}^- \widetilde \psi_-.
\end{align}
Here the covariant derivatives and the field strengths are defined by 
$D_i^\pm \ast = \partial_i \ast + [ A_i^\pm, \ast ]$,
${\cal D}_i U = \partial_i U + A_i^+ U - U A_i^-$
and 
$F_{ij}^\pm = \partial_i A_j^\pm - \partial_j A_i^\pm + 
[ A_i^\pm , A_j^\pm ]$. Let us substitute these expressions into the
D8-brane action:
\begin{align}
 S_{D8}
  =&\tilde{T}(2\pi\alpha')^2\int d^4x dz \frac{d\Omega_4}{V_4}~ 
  2\mbox{Tr}
  \left[\frac{R^3}{4U_z}\eta^{\mu \sigma}\eta^{\nu\tau}F_{\mu\nu}
 F_{\sigma \tau}
   +\frac{9U_z^3}{8U_{\rm KK}}\eta^{\mu\nu}F_{\mu z}F_{\nu z}
   \right.\nonumber\\
 &\left.\qquad
   +\frac{1}{2}\eta^{\mu\nu} h^{ij}F_{\mu i}F_{\nu j}
   +\frac{9U_z^4}{8R^3U_{\rm KK}} h^{ij}F_{zi}F_{zj}
   +\frac{U_z}{4R^3} h^{ik} h^{jl}F_{ij}F_{kl}
  \right].
\end{align}
Among the terms induced by the instanton, the $F_{\mu i}^2$ terms give 
kinetic terms and interactions, and the $F_{zi}^2$ and the $F_{ij}^2$
terms induce the pion mass and interactions. Substituting the expressions
of $F$ written in terms of $U$, we obtain the following action:
\begin{align}
 S_{D8}
 &=\tilde{T}(2\pi\alpha')^2 \int d^4x
  dz \frac{d\Omega_4}{V_4}~ 
 2{\rm Tr}
 \bigg[\frac{R^3}{4U_z} 
 [U^{-1}\partial_\mu U,U^{-1}\partial_\nu U ]^2
 \psi_+^2(\psi_+-1)^2  
 \notag \\[1mm]
 & \qquad
 +\frac{9U_z^3}{8U_{\rm KK}}(U^{-1}\partial_\mu U)^2 
 (\partial_z \psi_+)^2 
 \,+\,\frac{9U_z^4}{8R^3U_{\rm KK}}
 ( U^{-1} {\cal D}_i U )^2 ( \partial_z \widetilde \psi_+ )^2  
 \notag \\[1mm]
 &\qquad 
 + {1 \over 2}
 \Big(
 [U^{-1}\partial_\mu U, U^{-1} {\cal D}_i U ]
  \psi_+ \widetilde \psi_+ 
  + U^{-1} D_i^+ \big( \partial_\mu U U^{-1} \big) U \widetilde \psi_+
  - D_i^- \big( U^{-1} \partial_\mu U \big) \psi_+
 \Big)^2
 \notag \\
 &\qquad
 + 
\frac{U_z}{4R^3}
\Big(
 [U^{-1} {\cal D}_i U,U^{-1} {\cal D}_j U] 
 \widetilde \psi_+ (\widetilde \psi_+ - 1)
 +
 U^{-1} F_{ij}^+ U \widetilde \psi_+ 
 +
 F_{ij}^- \widetilde \psi_-
\Big)^2
\bigg].  \label{chiral L} 
\end{align}
Here the indices $\mu,\nu$ and $i$ are contracted by $\eta^{\mu \nu}$
and $h^{ij}$, respectively. 

Let us consider the expansion of the $U(x,\theta)$ field using the mode
functions for the case without the instanton. The lowest mode in such
expansion is just constant and thus
\begin{equation}
 U(x,\theta) 
 = 
 \exp 
 \big( 2 i \pi (x)/f_\pi + \textrm{higher $S^4$ KK modes} \big). 
 \label{U const}
\end{equation}
Here $\pi(x)=\pi^a(x)T^a$ is the pion field, which we choose 
to be Hermitian.
If we neglect the higher modes in this expression as 
$U=U(x)=\exp(2 i \pi(x) /f_\pi)$ and substitute it into \eqref{chiral L}, 
we have the following four-dimensional chiral Lagrangian:
\begin{align}
 \cal{L}
 &= {f_\pi^2 \over 4} {\rm Tr} (U^{-1}\partial_\mu U)^2 
 +C\;  \int \frac{d\Omega_4}{V_4}  {\rm Tr} 
\left(
U^{-1} A_i^+ U A_i^- 
\right) + {\cal O}(\mu^{-4}).
\label{chiral L2}
\end{align}
Here $f_\pi$ and $C$ are given by
\begin{align}
 &f_\pi^2
\equiv
\tilde{T}(2\pi\alpha')^2\int dz~ 
   \frac{9U_z^3}{U_{\rm KK}} \big( \partial_z \psi_+ \big)^2
   , \quad
C\equiv
   -\tilde T (2 \pi \alpha')^2 \int dz 
{9 U_z^4 \over 2 R^3 U_{\rm KK} } 
 (\partial_z \widetilde \psi_+)^2.
\end{align}
In (\ref{chiral L2}), we show terms of leading order in chiral
perturbation theory. 
In the previous subsection
we saw that the magnitude of the effect of the instanton with the size
$\mu$ can be estimated as $D_i \sim \epsilon \sim \mu^{-1}$  (see
\eqref{A_mu mass}). Hence we can treat
${\cal D}_i \sim D_i^\pm \sim \mu^{-1}$ and $F_{ij}^\pm \sim \mu^{-2}$. 
In addition, we took into account the fact that usually 
in chiral perturbation theories one counts the dimension of the
momentum $\partial_\mu$ in the same manner, 
$\partial_\mu \sim m_\pi\sim \mu^{-1}M_{\rm KK}$.

The pion mass term, i.e.~the second term in \eqref{chiral L2} does not
have the same form as the lowest mass term for pions in the chiral
perturbation with non-zero quark masses, $U\chi+\chi^\dag U^{-1}$ 
where $\chi$ is related to the bare quark masses.
The form \eqref{chiral L2} of the chiral Lagrangian suggests that our
deformation corresponds to turning on (an infinite number of)
external fields $A_i^\pm (\theta)$ in the QCD. From the transformation
laws under the chiral transformation in (\ref{baka}), we can read out
possible lowest terms of the perturbation to the QCD action:
\begin{eqnarray}
 {\cal L} = {\cal L}_{QCD}
  +G^{ap}_{bq}\bar{q}_{La}q_{R}^{~~q}\bar{q}_{Rp}q_{L}^{~~b}
  +h.c.,
\end{eqnarray}
where $a,b$ ($p,q$) are the indices of $SU(N_f)_L$ ($SU(N_f)_R$) and 
$q_L$ ($q_R)$ is the left-handed (the right-handed) quark field. 
The tensor $G^{ap}_{bq}$ is related to the two sources $A_i^+$ and
$A_i^-$. From the symmetry arguments we can only guess possible terms
with which the chiral symmetry is explicitly broken
and in particular
$A_i^\pm$ do not couple left- and right- handed quarks in a bilinear
manner.

When all the masses for the quarks are the same, the masses of the 
pions are equal to each other, because $SU(N_f)_V$ is unbroken. 
In our case, although both the sources
$A_i^+$ and $A_i^-$ break all the chiral symmetries explicitly because of
(\ref{baka}), the masses of the pions in \eqref{chiral L2} are the same.
This happens due to the fact that the BPST instanton has a symmetric
structure among the gauge indices $a$ and the spacetime coordinates $i$
and the contribution of intanton becomes gauge-blind after the 
contrcting the Lorentz indices $i$.

Before ending this subsection we will give some remarks. In the previous
subsection, we obtained the lowest $S^4$ mode in the presence of
instanton, $\zeta(\rho)$. If we use this function in expanding
$U(x,\theta)$ as:
\begin{equation}
 U(x,\theta) = 
 \exp \left( 2 i \pi (x) \zeta(\rho) / \tilde f_\pi + 
       \textrm{ higher $S^4$ KK modes } \right)
 \label{U zeta}  ,
\end{equation}
then even we neglect the higher $S^4$ KK modes the Lagrangian
\eqref{chiral L} do not reduce to the four-dimensional chiral Lagrangian
written in terms of $U$. Expanding the Lagrangian with using this $U$
we have:
\begin{equation}
{\cal L} =- {1 \over 2}  
  (\partial_\mu\pi^a)^2
  - {1 \over 2} m_\pi^2(\pi^a)^2
  + \frac{1}{4 \tilde e^2 \tilde f^2_\pi}
  ([\partial_\mu\pi,\partial_\nu\pi]^a)^2
  +\cdots. 
\end{equation}
Here the parameters are defined as follows:
\begin{align}
 \tilde f_\pi^2&\equiv\tilde{T}(2\pi\alpha')^2\int dz~ 
  \bigg\{
   \frac{9U_z^3}{U_{\rm KK}} \big( \partial_z \psi_+ \big)^2
   +
   4 \epsilon^2 \big( \tilde \psi_+ - \psi_+ \big)^2
   \bigg\}
   ,\\
   \frac{\tilde f_\pi^2}{2 \tilde e^2}&\equiv\tilde{T}(2\pi\alpha')^2\int 
  dz~ \frac{8R^3}{U_z}\psi_+^2(\psi_+-1)^2
  \int\frac{d\Omega_4}{V_4}\zeta^4,\\
  m^2_\pi &\equiv
 \epsilon^2 \frac{\tilde{T}(2\pi\alpha')^2}{\tilde f_\pi^2}
  \int dz~ \frac{9U_z^4}{R^3U_{\rm KK}}\big( \partial_z \tilde \psi_+ 
 \big)^2. 
 \label{chiral mass}
\end{align}
In deriving these, we have used the eigenvalue equation and the
normalization condition for $\zeta$ which we used in the previous
subsection. It can be checked that the pion mass \eqref{chiral mass}
reproduces the values which are consistent with those in the previous
subsection for large $\mu$.

The final point is related with the tachyon field. As discussed
in~\cite{Sugimoto:2004mh}, a tachyon field, which is the lowest mode of a
string stretching between D8 and $\overline{\rm D8}$,  couples with the 
quark bilinear, and if it develops a vacuum expectation value (VEV),
the quark mass terms will be generated. The tachyon field belongs to the
bifundamental representation of the chiral symmetry group 
$SU(N_f)_L \times SU(N_f)_R$. Inspired by this, let us add a complex
scalar field $T$ on the D8-branes which belongs to a fundamental
representation of the group of the gauge symmetry on the D8-branes,
and let it develop a VEV $\langle T\rangle$. Following the same procedure
the tachyon field 
will be expanded as
\begin{align}
 T(x,z,\theta) &= \xi_+(x,\theta)T_+ \Psi_+(z)+\xi_-(x,\theta)
 T_-\Psi_-(z),
 \hspace{3ex}  T_+=T_-=\langle T\rangle,
\end{align}
with 
non-normalizable
modes $\Psi_\pm$ 
(where $\Psi_-\equiv 1-\Psi_+$, $\Psi_+(z=\infty) =1$ and 
$\Psi_+(z=-\infty) =0$). 
The lowest
contribution to the chiral Lagrangian comes from the kinetic term
$g^{zz}|\partial_zT|^2$,
\begin{eqnarray}
  T_{\rm D8}\int d^9 \sigma e^{-\phi}\sqrt{-\det g}g^{zz}
  \mbox{Tr} |U^{-1}T_+-T_-|^2 \big(\partial_z \Psi_+ \big)^2 \propto
  \int d^4x~
  \mbox{Tr}[U\chi+\chi^{\dag}U^{-1}], \hspace{3ex}
\end{eqnarray}
with $\chi=T_- T_+^\dag$. This is exactly the expected form of the pion
mass term induced by the non-zero quark masses in the chiral perturbation
theory. The actual tachyon mass profile is complicated and should depend
on the spacetime coordinates in a curved geometry (see
\cite{Antonyan:2006vw} for a profile of fundamental strings in the
D4-brane geometry).

\section{Conclusion and discussion}

In this paper we have studied deformations of the holographic QCD model
considered by Sakai and Sugimoto, in order to introduce explicit chiral
symmetry breaking and non-zero pion mass. In the Sakai-Sugimoto model,
the chiral symmetry is realized by the D-brane configuration in the weak
coupling regime; flavor D8- and $\overline {\rm D8}$-branes are separated
from each other. So we have considered deformations of the D8- and 
$\overline {\rm D8}$-branes to connect them in the flat spacetime.

In section \ref{some attempts}, we have considered the configuration in
which the D8- and $\overline {\rm D8}$-branes are connected by a throat
and color D4-branes are placed in the throat. First we placed the
D4-branes at the center of the throat. Because the size of the throat 
is of the same order as the asymptotic distance between
the D8- and $\overline {\rm D8}$-branes, they do not reach the
near horizon region of the D4-brane background. Without taking the near
horizon limit, we have studied the spectrum of the gauge fields and we
still found a massless pion. Of course it is possible that without taking
the near horizon limit, this model just fails to capture the strong
dynamics of the dual QCD. In order to understand this point further, we
took a certain limit of this brane configuration in which the system
reduces to the simpler one; We placed the D4-branes close to the
D8-branes and magnified around a point on an angular $S^4$ of the
D8-brane worldvolume. Then the system reduces to the flat D4- and the
flat D8-branes located parallel to each other.
Because the D8-branes can be located arbitrarily close to the D4-branes, 
we can take the near horizon limit of the D4-brane background 
keeping the D8-branes in the near horizon region. In this case, we found
that the pions successfully acquire non-zero mass. However the minimum
value of the pion mass we obtained is too large; about $0.8$ times the
$\rho$ meson mass. This model has no parameter which can be tuned to take
the massless pion limit.

In order to construct a model which reproduces the small pion mass,
we considered a different deformation in section 
\ref{D4-brane charge and pion mass}. We introduced an instanton
background on the D8-branes considered by Sakai and Sugimoto. From the
point of view of the D-brane configuration, the instanton charge
corresponds to a smeared D4-brane charge, and the existence of this
charge assures that the D8- and the $\overline {\rm D8}$-branes are
connected to each other, even in the flat spacetime. Studying the
spectrum of the fluctuations, we found a non-zero pion mass which is
tunable using the parameter $\mu$, the size of the instanton, and
realized the observed value of the pion mass. We have found also that
the vector-meson spectrum is not affected. In section 
\ref{chiral perturbation}, we studied a chiral perturbation for QCD
which corresponds to introducing the instanton background. Using the mode
expansions without taking into account the perturbations induced by the
instanton, we have derived the four-dimensional chiral Lagrangian.
We also discussed the corresponding perturbations to the QCD Lagrangian.
A possible lowest candidate for the perturbation is a four-Fermi
coupling, not the quark mass term. Also the fact that the mass term of 
the Nambu-Goldstone boson for the axial $U(1)$ part has not been
generated by the instanton suggests that our deformation differs from
adding the non-zero quark mass. Using the lowest $S^4$ mode in the
presence of the instanton, we have obtained the pion masses which are
consistent with the values in the subsection \ref{meson spectrum}.

A final remark is a discussion on the throat solution in DBI action
and the quark mass. If we assume that the DBI action is reliable even 
for the small D8 throat considered in appendix 
\ref{Small throat solution},
then the possible quark mass estimated from the length of an open string
stretching between the D4- and D8-branes is finite. Although the D-brane
picture may break down for the thin D8 throat with the radius of the
string length, the throat can be thought as the result of partial tachyon
condensation~\cite{Hashimoto:2002xt}. The chiral symmetry is broken due
to the condensation and the quark mass terms are expected to be induced
since the tachyon couples with quarks bilinearly (see 
\cite{Casero:2007ae} for a recent
discussion along this direction). It is deserved to
understand more clearly the size of throat in the flat space in terms of
weak coupling physics and it is important to understand how to implement
this tachyon field into the probe D8-brane theory. We leave this issue as
a future problem.

\section*{Acknowledgements}

K.H.~would like to thank Yoshio Kikukawa, 
Horatiu Nastase, Tadakatsu Sakai, 
Shigeki Sugimoto and
Piljin Yi for useful discussions. 
T.H.~would like to thank the theory group of 
Institute of Physics, the University of Tokyo for the
hospitality during his visit.
A.M.~would like to thank Yoshio Kikukawa and Yoshihiro Mitsuka, 
for valuable discussions and comments.
K.H.~is partly supported by JSPS and 
the Japan Ministry of Education, Culture, Sports, Science and
Technology.
This work of T.H.~is partially supported by the
European Union 6th framework program MRTN-CT-2004-503069
``Quest for unification'', MRTN-CT-2004-005104 ``ForcesUniverse'',
MRTN-CT-2006-035863 ``UniverseNet'' and
SFB-Transregio 33 ``The Dark Universe'' by Deutsche
Forschungsgemeinschaft (DFG).
The work of A.M.~is supported in part by JSPS Research
Fellowships for Young Scientists. 
K.H.~thanks the Yukawa Institute for Theoretical Physics at Kyoto
University, for providing stimulating atmosphere for discussions
on this work, during the YITP workshops YITP-W-06-11 
``String Theory and Quantum Field Theory'', and YITP-W-06-16
``Topological Aspects of Quantum Field Theory''.

\appendix

\section{D8 probe in the D4 geometry}\label{hole}

In this appendix, we discuss the D8-brane probe solutions in the full
non-extremal D4-brane geometry \cite{Horowitz:1991cd} compactfied on an
$S^1$ without taking the near horizon limit. 
We show that there are a set of solutions which have the same
asymptotic behavior. The first solution is the one used by
Sakai and Sugimoto and the second solution corresponds 
to the configuration of D8- and $\overline{\rm D8}$- branes 
connected to each other by a throat.

The non-extremal D4-brane geometry is given by
\begin{align}
 & ds^2= \left(\frac{U^3}{R^3+U^3}\right)^{1/2}(dx_4^2+f(U)d\tau^2)
 +\left(\frac{U^3}{R^3+U^3}\right)^{-1/2}
 \left(\frac{dU^2}{f(U)}+U^2d\Omega_4^2
 \right) ,\\
 & e^{\phi}=\left(\frac{U^3}{R^3+U^3}\right)^{1/4}, \quad 
 F_4=\frac{2\pi N_c}{V_4}\epsilon_4, \quad
 f(U)=1-\frac{U_{\rm KK}^3}{U^3}.
\end{align}
This geometry is regular if and only if the compact $\tau$ direction has
a periodicity $\tau\sim \tau+\delta\tau$, 
$\delta\tau=4\pi R^{3/2}/3U^{1/2}_{\rm KK}$.
The near horizon geometry (\ref{kaba}) is obtained by taking the small
$U/R$ limit.

We place a probe D8-brane so that its configuration in $\tau$-$U$ plane
is given by $\tau=\tau(U)$ or equivalently by $U=U(\tau)$. With this
ansatz, the relevant part of the action is
\begin{align}
 S_{D8} =& -T_{D8}\int d^9 \sigma e^{-\phi}\sqrt{-\det g}
 \propto \int d\tau 
 U^4
 \sqrt{f(U)+\frac{R^3+U^3}{U^3}\frac{U'^2}{f(U)}} .
\end{align}
Since the action does not depend on $\tau$ explicitly, there is a
conserved quantity:
\begin{eqnarray}
 \frac{U^4 f(U)}{\sqrt{f(U)+\frac{R^3+U^3}{U^3}\frac{U'^2}{f(U)}}}
  =U_0^4\sqrt{f(U_0)},
\end{eqnarray}
where $U_0=U(\tau_0)\geq U_{\rm KK}$ is the point where
$U'(\tau_0)=0$. We can easily solve this equation and obtain
\begin{align}
 \tau(U) - \tau_0
 =&\int_{U_0}^UdU\frac{1}{f(U)}\sqrt{
  \frac{R^3+U^3}{U^3}\frac{U_0^8f(U_0)}{U^8f(U)-U_0^8f(U_0)} } .
\end{align}
It is easy to see that as $U$ goes to infinity $\tau(U)$ goes to a
constant which gives the asymptotic distance between the D8- and 
the $\overline{\rm D8}$-branes, 
$L=2\tau(U\rightarrow\infty)$ modulo $\delta \tau$.
When $U_0=U_{\rm KK}$, $L=\delta\tau/2$ and this is the solution
considered in \cite{Sakai:2004cn}. As the difference $U_0-U_{\rm KK}$
gets larger, $L$ first decreases, but around a critical value 
$U_0-U_{\rm KK}\sim 7$ (for $R=10$ and $U_{KK}=1$), $L$ starts increasing
and it continues to increase after passing this value. So, in fact, we
found two solutions for fixed $\delta\tau$. For example, for
$L=\delta\tau/2$, $U_0=U_{\rm KK}$ is an obvious
solution found before, and the second solution with
$U_0\sim U_{\rm KK}+70$ for $R=10$ and $U_{KK}=1$
corresponds to the configuration of the D8- and $\overline{\rm D8}$-
branes connected with each other by a throat. (The other solutions with
$L=\delta\tau/2$ have $2\tau(U\rightarrow\infty)>\delta\tau$  
in the covering space of the $S^1$,
and thus these solutions correspond to the probe D8-branes
wrapping more than once around the $S^1$ direction.) Since the ratio
$U/R$ is small in the near horizon region, the second solution exists
outside of the near horizon region, and thus it disappears when we take
the near horizon limit.

We can now study whether the pions become massive. Although we do not
take the near horizon limit, we simply apply the AdS/CFT correspondence
for studying if we have Nambu-Goldstone bosons. We check it by studying
the normalizability of the zero mode in the $U$-direction (or
$z$-direction in the main text) of the gauge fields on the D8-brane since
this mode would be identified as the pion. If it is normalizable, the
zero mode is the pion. If it is not instead, there is no Nambu-Goldstone
boson. In the latter case, the situation may be consistent with the pion
being massive because of the explicit chiral symmetry breaking. 

The relevant part of the action for the fluctuations of the gauge fields
on the D8-brane is
\begin{align}
 S =&-T_{D8}(2\pi\alpha')^2\int d^9 \sigma
 e^{-\phi}\sqrt{- \det g}
 {1 \over 4}g^{MN} g^{PQ} F_{MP} F_{PQ} \notag
  \\
 \sim&-\int d^4x dU U^4 \left(\frac{U^3}{R^3+U^3}\right)^{-1/2}
 \sqrt{\frac{U^8}{U^8f(U)-U_0^8f(U_0)}} \notag\\
 & \times \left[\left(\frac{U^3}{R^3+U^3}\right)^{-1}
   \eta^{\mu\rho}\eta^{\nu\sigma}F_{\mu\nu}F_{\rho\sigma}
    +
     2 \frac{U^8f(U)-U_0^8f(U_0)}{U^8} 
    \eta^{\mu\nu}F_{\mu U}F_{\nu U}
   \right] ,
\end{align}
where $\eta^{\mu\nu}$ is the four-dimensional Minkowski metric.
Then the zero mode in $A_U$ is given by
\begin{equation}
 A_U(x,U)= \pi(x)\phi_0(U), \hspace{3ex}
  \phi_0(U)\propto 
  \left(\frac{U^3}{R^3+U^3}\right)^{1/2}
  \frac{1}{\sqrt{U^8f(U)-U_0^8f(U_0)}},
\end{equation}
and it is easy to see that the zero mode is normalizable:
\begin{equation}
 \int_{U_0}^\infty dU 
  \left(\frac{U^3}{R^3+U^3}\right)^{-1/2}
  \sqrt{U^8f(U)-U_0^8f(U_0)}
  \phi_0(U)^2 <\infty .
\end{equation}
Thus we always have a massless Nambu-Goldstone boson for arbitrary $U_0$.
Therefore even though the D8- and $\overline{\rm D8}$-branes are
connected by a throat in the flat spacetime, the pion is still massless.

\section{D8-brane parallel to D4-brane}\label{pp}

In this appendix, we compute the pion mass in the D-brane configuration
in which a D8-brane is placed parallel to the color D4-branes. The
effective theory is a QCD with massive quarks. Although the masses
of the quarks are roughly of the same order as the compactification
scale, we may still expect that the pion masses are suppressed compared
with the other meson masses. 

For our later convenience, we introduce a new coordinate $r$ defined by
\begin{equation}
  r=\Bigg(\frac{\sqrt{U^3}+\sqrt{U^3-U_{\rm KK}^3}}{2}\Bigg)^{2/3},
   \qquad
  U=\left(r^{3/2}+\frac{U_{\rm KK}^3}{4r^{3/2}}\right)^{2/3},
\end{equation}
and the near horizon limit of the D4-brane geometry is
\begin{align}
 ds^2=& \left(\frac{U}{R}\right)^{3/2}(dx_4^2+f(U)d\tau^2)
 +\left(\frac{R}{U}\right)^{3/2}\frac{U^2}{r^2}dX_5^2 ,\\
 dX_5^2=&dy^2+y^2d\Omega_3^2+dw^2,\hspace{3ex} r^2=y^2+w^2 .
\end{align}
We introduce the D8-brane in such a way that it is parallel to the
D4-branes and thus 
it is localized in the $w$-direction. The position
in the $w$-direction becomes a function of the $y$ coordinate,
$w=w(y)$. Then the induced metric and the equation of motion for $w(y)$
computed from Dirac-Born-Infeld action for the D8-brane probe are given
by
\begin{align}
 ds^2=& \left(\frac{U}{R}\right)^{3/2}(dx_4^2+f(U)d\tau^2)
 +\left(\frac{R}{U}\right)^{3/2}\frac{U^2}{r^2}
 (dy^2+y^2d\Omega_3^2+w'(y)^2dy^2),
 \\
  w''(y)=& (1+w'(y)^2)\left(
  -\frac{3}{y}w'(y)
  +\frac{3U_{\rm KK}^3}{4r^5}\left(w(y)-y w'(y)\right)
  \left(\frac{1}{1-\frac{U_{\rm KK}^3}{4r^3}}
  -\frac{5}{3}\frac{1}{1+\frac{U_{\rm KK}^3}{4r^3}}\right)
  \right) .
\end{align}
The solution $w(y)$ of this equation satisfies the asymptotic behavior
\begin{eqnarray}
 w(y) \sim m +\frac{\nu(m)}{y^2},
\end{eqnarray}
where $\nu(m)$ is determined so that the solution $w(y)$ is regular
everywhere, and $m$ is a parameter which describes the asymptotic
distance between the D4- and D8-branes. Since the quark masses receive
corrections through integrating out massive Kaluza-Klein modes along 
the $\tau$ direction, there is no identification that $m$ is the mass of
the quark (at the compactification scale) and $\nu(m)$ is proportional to
the value of the chiral condensate.

The fluctuations in $A_\mu$ on the D8-branes are identified as vector
mesons in 
QCD and so the lowest mode is identified as the $\rho$
meson. We can safely take the $A_y=0$ gauge since there is no scalar zero
mode among the fluctuations of $A_\mu$. Then the lowest Kaluza-Klein mode
in the fluctuation $A_\tau$ can be identified as the massive pion. The
action for fluctuations on the D8-brane at the quadratic level is given
by
\begin{align}
 S=&-T_{D8}(2\pi\alpha')^2\int d^9 \sigma e^{-\phi}\sqrt{-\det g} 
 {1 \over 4} g^{MN} g^{PQ} F_{MP} F_{NQ} \notag 
  \\
  \propto&-
  \int dy \left(1-\frac{U_{\rm KK}^3}{4r^3}\right)
  \left(1+\frac{U_{\rm KK}^3}{4r^3}\right)^{5/3}y^3\sqrt{1+w'^2}
  \left\{
  \left(\frac{R}{U}\right)^3
  \eta^{\mu\rho}\eta^{\nu\tau}F_{\mu\nu}F_{\rho\tau}
  \right.
  \nonumber\\
  &\qquad
  \left.
  +\frac{2r^2}{U^2(1+w'^2)}\eta^{\mu\nu}F_{\mu y}F_{\nu y}
  +\frac{2R^3}{U^3f(U)}\eta^{\mu\nu}F_{\mu\tau}F_{\nu\tau}
  +\frac{2r^2}{f(U)U^2(1+w'^2)}F_{y \tau}^2
  \right\},
\end{align}
where we have considered only the $S^3$-independent modes and omitted the
$S^3$ components of the gauge fields. Performing the fluctuation analysis
we have numerically obtained the result that the mass ratio
$m_\pi/m_\rho$ is about $0.8$. In the numerical calculation, we have
chosen the asymptotic distance between the D4- and the D8-branes such
that the D8-brane is located close to the D4-branes, but does not reach
the point $U=U_{\rm KK}$. The reason for this choice is the following;
If the D8-brane is far away from the D4-branes the bare quark mass is
large, on the other hand if the D8-brane intersects with the D4-branes, 
a large correction to the quark mass is expected. In fact the ratio
$m_\pi/m_\rho$ becomes the smallest value for this choice. Therefore the
pion mass is much larger than the actual value in the real QCD.

\section{Mixing between lowest $S^4$ KK modes of $A_i$ with those of 
$A_{\nu}$ and $A_z$}  
\label{mixing with A_i}

In this appendix, we discuss the mixing terms between the lowest $S^4$ KK 
modes of the fluctuations $A_\nu$ and $A_z$ with those of $A_i$ in the
case with the instanton background.

The following mixing terms arise from the terms $F_{\nu i}^2$ and 
$F_{z i}^2$ in the Lagrangian:
\begin{equation}
 F_{\nu i}^2:\quad
 \int d^4X L(\rho) {\rm Tr} D_\alpha A_\nu \partial_\nu A_\alpha,
 \qquad
 F_{zi}^2:\quad 
 \int d^4X L(\rho) {\rm Tr} D_\alpha A_z \partial_z A_\alpha,
\end{equation}
where $D_{\alpha} A=\partial_{\alpha} A +[A_{\alpha}^{\rm inst},A]$ 
and $L(\rho)=4(\rho^2+1)^{-2}$. 
The index $\alpha=1,\ldots,4$ is that for the coordinates $X^\alpha$
of the $\textrm{\bf R}^4$ introduced in subsection \ref{instanton on D8}
and $\rho = |X|$ as before. The lowest modes of the fluctuations
$A_\alpha$ are given by the moduli $G_\alpha^{{\rm tr} \beta}$(X) and 
$G_\alpha^{\rm size}$(X) of the instanton which are the translation of
the instanton and the change of the instanton size
respectively.\footnote{
The other moduli corresponding to gauge directions can be
absorbed by appropriate gauge transformations and field redefinitions.
} 
Using these, the lowest modes of $A_\alpha$ are given by
\begin{align}
 A_\alpha(x,z,X) =& f_\beta(x,z) 
 G^{{\rm tr} \beta}_\alpha(X)
 +
  f_{\rm size}(x,z) 
 G^{\rm size}_\alpha(X).
\end{align}
The explicit forms of the translational moduli 
$G^{{\rm tr} \beta}_\alpha$ 
and the size modulus $G^{\rm size}_\alpha$ are as follows:
\begin{align}
 &
  G^{{\rm tr} \beta}_\alpha(X)
 \equiv
 {\partial \over \partial X^\beta} A^{\rm inst}_\alpha (X) 
  =i { -\epsilon_{\alpha \beta c} \sigma_c 
 +\delta_{\alpha 4} \sigma_\beta 
 -\delta_{\beta 4} \sigma_\alpha \over \mu^2 + \rho^2}
 -{2 X^\beta \over \mu^2 + \rho^2} A^{\rm inst}_\alpha (X), 
 \label{moduli_1}\\
 & 
 G^{\rm size}_\alpha(X)
 \equiv
 {\partial \over \partial \mu} A^{\rm inst}_\alpha (X) 
 = -{2 \mu \over \mu^2 + \rho^2} A^{\rm inst}_\alpha (X).
 \label{moduli_2}
\end{align}
Here the parameter $\mu$ controls the size of the instanton. 
The totally antisymmetric tensor $\epsilon_{\alpha\beta\gamma}$
is defined by $\epsilon_{123}=1$ and $\epsilon_{4 \alpha \beta}=0$
and $\sigma_4=0$. Note that the translational moduli 
$G^{{\rm tr} \beta}_\alpha(X)$ are even under the four-dimensional parity
transformation $\{ X^\gamma \} \rightarrow \{ -X^\gamma\}$. Then the
mixing term in $F_{\nu i}^2$ which includes $\partial_{\nu} f_\beta(x,z)$
becomes
\begin{align}
\int d^4X L(\rho)
 {\rm Tr}
 \bigg[
 G^{{\rm tr}\beta}_\alpha(X) 
 \bigg\{ \tilde{A}_\nu(x,z)
 \frac{X^\alpha}{\rho}\partial_\rho \zeta(\rho)
 +\big[A_\alpha^{\rm inst},\tilde{A}_\nu(x,z)\zeta(\rho)\big]
 \bigg\} \bigg],
\end{align}
where the lowest fluctuation modes of $A_\nu$ are given by 
$A_\nu(x,z,X)=\tilde{A}_\nu(x,z)\zeta(\rho)$ as in subsection 
\ref{meson spectrum}. Since the inside of the square bracket is odd under
the parity transformation $\{X^\gamma\} \rightarrow \{-X^\gamma\}$, this
mixing term vanishes after the integration over the ${\bf R}^4$. On the
other hand, the mixing term in $F_{\nu i}^2$ which includes
$\partial_\nu f_{\rm size} (x,z)$ becomes
\begin{align}
 \int d^4X L(\rho)
 {\rm Tr}
 \bigg[
 G^{\rm size}_\alpha (X)
 \bigg\{\tilde{A}_\nu(x,z)\frac{X^\alpha}{\rho}\partial_\rho \zeta(\rho)
 +\big[A_\alpha^{\rm inst},\tilde{A}_\nu(x,z)\zeta(\rho)\big]
 \bigg\}
 \bigg]
 ,
\end{align}
and it vanishes since the size moludus $G^{\rm size}_\alpha (X)$ 
is proportional to the instanton $A_\alpha^{\rm inst}$. 
Similarly we can also show that the mixing terms from $F_{zi}$ vanish.
Thus there is no mixing between the lowest $S^4$ KK modes of the 
fluctuations $A_\nu$ and $A_z$ with those of $A_i$.

\section{Small throat solution in Dirac-Born-Infeld action} 
\label{Small throat solution}

In this appendix, we estimate the throat radius of a classical solution
of the D8-brane effective action in the flat spacetime background, 
when the D4-brane charge (the instanton on the $S^4$) is introduced. 
This would correspond to computing the quark mass, from the length of the
string stretching between the D8 throat surface and the $N_c$ D4-branes.

Our situation is similar to the one in \cite{Constable:2001ag} where 
a set of $N$ D1-branes ending on $n$ D5-branes is obtained as a
deformation of the 
surface of the D5-branes. We take T-dualities three times along the
transverse directions, and obtain our brane configuration of the $N$
D4-branes ending on the $n$ D8-branes. In this paper we took $N=1$ and
$n=2$ (one instanton on the $S^4$ in the SU(2) Yang-Mills). 

The difference from \cite{Constable:2001ag} is only that
\cite{Constable:2001ag} treats infinitely long 
D4-branes
while in our
case the D4-brane is ending on the $\overline{\rm D8}$-branes. 
In \cite{Callan:1997kz}, two throat solutions connecting the D8- and the
$\overline{\rm D8}$-branes are obtained, one has the throat whose radius
is almost equal to the D8-$\overline{\rm D8}$ distance, while the other 
has a
small radius of the size of string length. What we are
interested in is the latter one, which exists only when one introduces
the D4-brane charge on the D8-branes (in \cite{Callan:1997kz}
fundamental string charge is introduced, instead). Let us construct this
latter type of the solution in our situation.

Following (43) and (45) of \cite{Constable:2001ag}, the equation
of motion for the scalar field $\tau$ of the D8-branes in the flat
background spacetime is written as 
\begin{eqnarray}
\frac{d\tau/dU}{\sqrt{1 + (d\tau/dU)^2}} 
= \frac{(2\pi\alpha')^{2}/\tilde{B}}{nU^4 + (3N/2)(2\pi\alpha')^2}.
\end{eqnarray}
Here we assumed that the $N$ 
instantons on the D8-brane surface 
are
homogeneous on the $S^4$ (corresponding to $\mu=1$ in this paper).
Note that we use the same $U$ as the radial coordinate of the D8-branes, 
{\it in the flat spacetime}. $\tilde{B}$ is an integration constant.
The throat radius $U_{\rm throat}$ can be defined where 
the derivative $d\tau/dU$ diverges, that means that the above equation is
equal to the unity. 

The distance between the D8-branes and the $\overline{\rm D8}$-branes
is in our case half the circumference of the $S^1$, $\delta\tau/2$. 
Integrating the above equation, we obtain 
\begin{eqnarray}
\frac{\delta\tau}{2} = 2\int_{U_{\rm throat}}^\infty \!\!\! dU \;
\left(
\tilde{B}^2 \left(
nU^4/(2\pi\alpha')^{2} + 3N/2
\right)^2-1
\right)^{-1/2}.
\end{eqnarray}
This integral equation determines the relation between $U_{\rm throat}$
and the integration constant $\widetilde{B}$. We evaluate this in the low
energy limit $\alpha' \to 0$ while $\delta \tau$ fixed. This is in fact
the limit similar to the one taken in \cite{Callan:1997kz}, and we apply
their choice of the integration constant so that we have the small radius
of the throat:
\begin{eqnarray}
 \tilde{B} =\frac{2}{3N}(1-\epsilon), \quad 
\epsilon \to +0 (\propto \alpha'^2).
\end{eqnarray}
Then, up to overall numerical constants, we obtain
\begin{eqnarray}
 U_{\rm throat} \sim (\alpha')^{1/2} \epsilon^{1/4}, \quad
\delta\tau \sim (\alpha')^{1/2}\epsilon^{-1/4}.
\end{eqnarray}
Therefore for a fixed $\delta\tau$ and 
in the low energy limit $\alpha'\to 0$, 
the throat radius is estimated as
\begin{eqnarray}
 U_{\rm throat} \sim \alpha'/\delta\tau.
\end{eqnarray}
This provides the naive evaluation of the energy of the lowest excitation
of the string stretched between the throat surface and the throat center
(the $N_c$ D4-branes), which is expected to be the quark mass, as
\begin{eqnarray}
 m_{\rm quark} \sim U_{\rm throat}\alpha^{'-1}
\sim 1/\delta\tau. \label{qm}
\end{eqnarray}
Here the string tension is proportional to $\alpha^{'-1}$. We find that
the quark mass is finite in the low energy limit $\alpha'\to 0$.

In order to see whether the Gell-Mann--Oakes--Renner (GOR) relation
$m_{\pi}^2 =Bm_q$ (where $B$ is related to the quark condensate) is
satisfied, let us evaluate the $\mu$ dependence of the quark mass
from this thin throat picture quantitatively. When we deform the
instanton distribution away from the homogeneous one, it is expected that 
the spherical cross-section $S^4$ is deformed in the analysis in the flat
spacetime background.\footnote{
In the strong coupling regime, the shape of the $S^4$ is deformed,
but this is an effect of higher order in $\alpha'$ and thus we have 
neglected it. The shape of the D8-brane is non-singular 
even in the $\alpha'\to 0$ limit.
On the other hand, in the weak coupling regime (the flat spacetime),
the shape becomes singular in the $\alpha'\to 0$ 
limit (the angular $S^4$ shrinks), thus the deformation should be
taken into account from the first place.
}
Thus the effective length of the stretched string is expected to be 
shortened by this deformation, and the quark mass may become lighter. 
To see this concretely, we consider the case with two instantons on the
$S^4$; one on the north pole and the other on the south pole, in order 
to maintain the parity symmetry $\theta^1\leftrightarrow \pi-\theta^1$
on the deformed shape of the $S^4$, for simplicity. The size of each
instanton is given by $1/\mu$. 
For large $\mu$, the instanton density at the poles is $F^2\sim \mu^4$
while that around the equator is $F^2\sim 1/\mu^4$. To maintain the
force balance on the $S^4$, the D8-brane shape should be deformed
in such a way that the outbound force generated by this instanton
density is canceled by the inbound force given by 
the D8-brane energy density per unit angular volume on the deformed
$S^4$. Due to the parity symmetry, we deform the $S^4$ to an 
ellipsoid whose major axis is generated by a multiplication by a factor
$\alpha$ while its minor is by a factor $\beta$. Then, to keep the
tension balance, one would need to require 
$\alpha \sim \mu$ and $\beta \sim 1/\mu$, because of the instanton
energy density. The minor axis has the
length proportional to $1/\mu$, so the quark mass is expected to scale
as $m_q \sim 1/(\mu\delta\tau)$. We find that this scaling is different
from what we obtained in this paper, 
$m_\pi \propto \epsilon \propto 1/\mu$, if we apply the GOR relation.

Our estimation of the deformation of the $S^4$  would have been too
naive. Furthermore, we have assumed that the quark mass can be given 
just by the shortest radius of the throat, which would be incorrect,
since the D8-brane surface is highly curved.
One should note that this DBI approximation is not a good low energy 
approximation, especially around the throat. It is possible that the 
quark mass may not be generated from the throat at all, which is
indicated in our results of the chiral perturbation.

\end{document}